\documentclass[a4paper,twoside]{article}

\usepackage{epsfig}
\usepackage{subfig}
\usepackage{calc}
\usepackage{amssymb}
\usepackage{amstext}
\usepackage{amsmath}
\usepackage{amsthm}
\usepackage{multicol}
\usepackage{pslatex}
\usepackage{apalike}
\usepackage{algorithm2e}
\usepackage[bottom]{footmisc}
\usepackage{SCITEPRESS}
\usepackage{url}
\usepackage{todonotes}
\usepackage{physics}
\usepackage{tikz}
\usetikzlibrary{quantikz2}
\usepackage[noend]{algpseudocode}
\usepackage{pgfplots}
\usepackage[outline]{contour}
\contourlength{1.2pt}
\usepackage{placeins}
\pgfplotsset{compat=1.18}
\usepackage{adjustbox}
\usepackage[hidelinks]{hyperref}

\begin{document}
\title{Towards Efficient Quantum Anomaly Detection: One-Class SVMs using Variable Subsampling and Randomized Measurements}

\author{\authorname{Michael Kölle\sup{1}, Afrae Ahouzi\sup{1,2}, Pascal Debus\sup{2}, Robert Müller\sup{1}, Danielle Schuman\sup{1} and Claudia Linnhoff-Popien\sup{1}}
\affiliation{\sup{1}Institute of Informatics, LMU Munich, Munich, Germany}
\affiliation{\sup{2}Fraunhofer AISEC, Garching, Germany}
\email{michael.koelle@ifi.lmu.de}
}


\abstract{
Quantum computing, with its potential to enhance various machine learning tasks, allows significant advancements in kernel calculation and model precision. Utilizing the one-class Support Vector Machine alongside a quantum kernel, known for its classically challenging representational capacity, notable improvements in average precision compared to classical counterparts were observed in previous studies. Conventional calculations of these kernels, however, present a quadratic time complexity concerning data size, posing challenges in practical applications. To mitigate this, we explore two distinct approaches: utilizing randomized measurements to evaluate the quantum kernel and implementing the variable subsampling ensemble method, both targeting linear time complexity. Experimental results demonstrate a substantial reduction in training and inference times by up to 95\% and 25\% respectively, employing these methods. Although unstable, the average precision of randomized measurements discernibly surpasses that of the classical Radial Basis Function kernel, suggesting a promising direction for further research in scalable, efficient quantum computing applications in machine learning.
}

\onecolumn \maketitle \normalsize \setcounter{footnote}{0} \vfill

\section{\uppercase{Introduction}} \label{sec:introduction}
In the prevailing digital landscape, anomaly detection algorithms safeguard numerous systems by identifying irregularities, such as unauthorized intrusion into a network, unexpected machinery behavior, or abnormal medical readings \cite{fernando2021deep,9377017}. In e-commerce, these algorithms are pivotal in protecting both customers and companies from fraudulent transactions, which could precipitate substantial financial losses \cite{EuropeanCentralBank_2023}. Anomaly detection faces challenges, including the diverse and rarely similar nature of anomalies, and often encounters high-dimensional, correlated, and sometimes unlabeled data sets. 

Quantum Machine Learning (QML), a synthesis of machine learning and quantum computing, promises potential solutions to some of these challenges by leveraging the capability of quantum algorithms to compute classically challenging kernels \cite{havlivcek2019supervised}. Some initiatives have aimed at ameliorating existing anomaly detection methods with quantum methods. A notable attempt \cite{kyriienko2022unsupervised} achieved a 20\% elevation in average precision using one-class support vector machine models with a quantum kernel, though confronted quadratic scaling challenges with data size, impacting both training and testing times.

Addressing the aforementioned time complexity challenge, this work replicates the results from \cite{kyriienko2022unsupervised} and employs them as benchmarks to explore the efficacy of two linear time complexity methods based on data size: \emph{randomized measurements} for quantum kernel measurement and an ensemble method termed \emph{variable subsampling}. The focus rests on two pivotal questions:
\begin{enumerate}
    \item Can the quantum kernel extract superior information from the data compared to the classical Radial Basis Function kernel, and thereby offer overall enhanced results?
    \item Do randomized measurements and variable subsampling models maintain the performance of the quantum kernel derived through the inversion test, while diminishing the time complexity related to data size?
\end{enumerate}
The evaluation engages synthetic data and the Credit Card Fraud data set, inspecting performance and time complexities contingent on data size and qubit number. Our findings, diverging from \cite{kyriienko2022unsupervised}, reveal that while attainable improvements in average precision and F1\_score over the classical kernel are discernible, they are minimally significant. Models utilizing variable subsampling with the inversion test exhibited stability, whereas those employing the randomized measurement method presented high variance. Nevertheless, variable subsampling did manifest considerable enhancements in training and testing times, indicating potential performance elevation opportunities through alternate hyperparameters.

\section{\uppercase{Preliminaries}} \label{sec:preliminaries}

\subsection{One-Class Support Vector Machines}\label{subsec:OC-SVM}

\begin{figure}[tb]
    \centering
        \begin{tikzpicture}[
        declare function={c(\x)=-\x+5;},
        declare function={a(\x)=-sqrt(0.4-(\x)^2);},
        declare function={b(\x)=sqrt(0.4-(\x)^2);},
        declare function={f(\x) = 2*sqrt(2)*rad(atan(\x/(2*sqrt(2))))*5/2.99;}]
        \begin{axis}[width=0.64*\linewidth,
            xmin=-2,
            xmax=2,
            ymin=-1.26,
            ymax=2,
            axis lines=middle,
            axis equal image,
            xtick=\empty, ytick=\empty,
            enlargelimits=true,
            clip mode=individual, clip=false,
            xlabel={\tiny$d_1$},
            ylabel={\tiny$d_2$},
            xlabel style={at={(ticklabel* cs:1)},anchor=north},
            ylabel style={at={(ticklabel* cs:1)},anchor=south}
        ]
        \addplot[only marks, mark={x}, samples=120]
            (f(x)+1, {0.5*(a(x)+b(x)) + rand * ( a(f(x)) - b(f(x))) / 2}+1.5);
        \addplot[only marks, mark={x}, samples=120]
            (-f(x)-1, -{0.5*(a(x)+b(x)) + rand * ( a(f(x)) - b(f(x))) / 2}-1.5);
        \addplot[red, thick, only marks, mark={+}, samples=4] (rand+rand, rand+rand);
        \node[below] at (current bounding box.south){\small Input space};
        \end{axis}
        \begin{axis}[width=0.7*\linewidth, 
        at={(0.6*\linewidth,0)},
        domain=-3:5,
        ymin=-4.5,
        axis lines=middle,
        axis equal image,
        xtick=\empty, 
        ytick=\empty,
        clip mode=individual, 
        clip=false,
        xlabel={\tiny$d'_1$},
        ylabel={\tiny$d'_2$},
        xlabel style={at={(ticklabel* cs:1)},anchor=north west},
        ylabel style={at={(ticklabel* cs:1)},anchor=south west}]
        \addplot[only marks, mark={x}, samples=10,domain=0.5:5](\x+rnd,{2*rnd-\x+5.5});
        \node[red, thick, inner sep=-2pt] at (-2.65,-0.1) (A1) {\tiny$+$};
        \node[red, thick, inner sep=-2pt] at (-2.5,2.75) (A2) {\tiny$+$};
        \draw[gray] (A1) -- ($(5,0)!(A1)!(0,5)$) node[midway,sloped,below,rotate=360, scale=0.8]{\contour{white}{\tiny$b-\mathbf{w}\cdot \Phi(x_i)$}};
        \draw[gray] (A2) -- ($(5,0)!(A2)!(0,5)$) node[near start,sloped,below,rotate=360]{};
        \draw[gray, <->, thick] (0,0,0) -- ($(5,0)!(0,0,0)!(0,5)$) node[midway,sloped,below,rotate=360]{\contour{white}{\tiny $b$}};
        \addplot[thick,domain=-3:8]{c(x)} node[below,very near start,sloped,rotate=360, scale=0.85]{\contour{white}{\tiny Separating hyperplane}} node[gray, below, very near end, sloped, scale=0.85] {\contour{white}{\tiny Anomalous}} node[gray, above, very near end, sloped, scale=0.85] {\contour{white}{\tiny Normal}};
        \node[below] at ($(current bounding box.south) + (-3,-5.8)$) {\small Feature space};
        \end{axis}
        \draw[->, gray](0.45*\linewidth,1.1) -- (0.55*\linewidth,1.1) node[gray, midway, align=center,text width=2.33cm]{\tiny Feature Map\\ $\Phi$};
    \end{tikzpicture}
    \caption{The linearly inseparable points on the input space are mapped using a quantum feature $\Phi$ into a feature space where they are linearly separable \cite{aggarwal2017introduction}.}
    \label{fig:margin-hyperplane}
\end{figure}
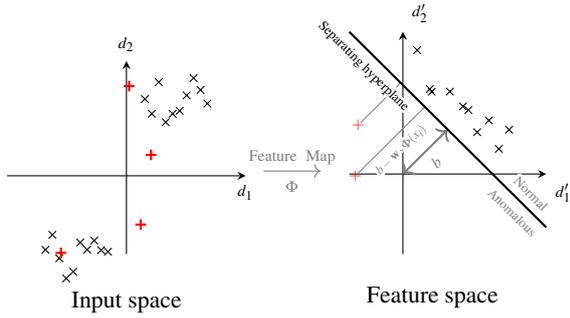

The One-Class Support Vector Machine (OC-SVM), originally presented by \cite{scholkopf1999support}, offers an unsupervised variant of the SVM for anomaly and novelty detection. Differing from the conventional SVM, which finds a maximum margin hyperplane to distinguish normal from anomalous data, the OC-SVM presupposes that the origin represents the anomalous class when labels are absent. It seeks to maximize the margin $b$ between origin and the input data, penalizing points below the hyperplane (\autoref{fig:margin-hyperplane}). These are governed by a hyperparameter \(\nu \in (0,1]\) which dictates the portion of points categorized as anomalous.

\begin{align}
    \min_{\mathbf{w}, b}\frac{1}{2} \mathbf{w}^2 + \frac 1 {\nu N} \sum_{i=1}^N \max\{b-\mathbf{w}\cdot \Phi(x_i),0\} - b \label{eq:objective-function}
\end{align}

To address non-linearly separable data, feature maps \(\Phi: \mathcal X \rightarrow \mathcal F\) elevate data from input space \(\mathcal X\) to a higher dimensional feature space \(\mathcal F\). Since direct calculations of feature maps are computationally intensive, a kernel function \(k: \mathcal X \times \mathcal{X} \rightarrow \mathbb{R}\) is used to ascertain data point similarity in the embedded feature space.
\begin{align}
    k(x_i, x_j) = \langle \Phi(x_i), \Phi(x_j) \rangle
\end{align}

The use of the kernel function, which can be summed up in matrix form by the kernel matrix \(G = [k(x_i,x_j)]\), facilitates a dual problem formulation. Solving this dual problem leads to an implicit parameterization of the hyperplane based on the support vectors $\alpha_i$. 

Relying on this parameterization, scoring new data depends on its relation to the separating hyperplane:
\begin{align}
    \text{Score}(x_\text{new}) = \sum_{i=1}^N \alpha_i \cdot k(x_{\text{new}}, x_i) \label{eq:decision-function}
\end{align}

The label is determined by the score's sign: negative connoting an anomaly and positive signifying normalcy.

\subsection{Quantum Kernel Embedding}\label{subsec:quantum_embedding}

\begin{figure*}[t]
    \centering
    \raisebox{0.7\height}{
    \subfloat[IQP-like feature map]{%
        \begin{adjustbox}{width=0.75\linewidth}
        \begin{quantikz}[transparent,font=\large]
        \ket{0} \ & \gate{H}\gategroup[5,steps=7,style={dashed,inner xsep=2pt}, label style={label position=above left}]{{\quad\quad\quad\quad\quad\quad\quad\quad\quad\quad\quad\quad\Large$\times 2\lambda$}} & \gate{R_Z(x_1)} & \gate[2]{R_{ZZ}(x_1 x_2)} &  & \gate[3,label style={yshift=0.3cm}]{R_{ZZ}(x_1 x_3)} &  \ \ldots\ & \gate[5,label style={yshift=0.3cm}]{R_{ZZ}(x_1 x_d)} &  & \rstick[5]{$U_\Phi(\mathbf{x})\ket{0}$}\\
        \ket{0} \ & \gate{H} & \gate{R_Z(x_2)} &  & \gate[2]{R_{ZZ}(x_2 x_3)} & \linethrough & \ \ldots\ & \linethrough &  & \\
        \ket{0} \ & \gate{H} & \gate{R_Z(x_3)} &  &  &  & \ \ldots\ & \linethrough &  & \\
        & \setwiretype{n}\vdots & & & & & \vdots & & & \\
        \ket{0} \ & \gate{H} & \gate{R_Z(x_d)} &  &  &  & \ \ldots\ &   &  & 
        \end{quantikz}
        \end{adjustbox}
        \label{fig:IQP-FeatureMap}
    }
    }
    \hfill 
    \begin{minipage}[b]{0.23\linewidth}
        \subfloat[Inversion test circuit]{%
            \begin{adjustbox}{width=\linewidth}
                \begin{quantikz}[transparent,font=\large]
                \ket{0} & \gate[4]{U_\Phi(x)} & \gate[4]{U^\dagger_\Phi(x')}&\meter{} &\\
                \ket{0} &  & &\meter{} &\\
                \setwiretype{n}\vdots &  & &\vdots &\\
                \ket{0} &  & &\meter{} &
                \end{quantikz}
            \end{adjustbox}
            \label{fig:Inversion-Test}
        }
        
        \vfill 
        
        \subfloat[Swap test circuit]{%
            \begin{adjustbox}{width=\linewidth}
            \begin{quantikz}[transparent,font=\large]
            \ket{0} \ \ \ & \gate{H} & \ctrl{2} & \gate{H} & \meter{} &  \\
            \ket{0}^{\otimes d}& \gate{U_\Phi(x)}\qwbundle{} & \swap{1} &  &  &  &\\
            \ket{0}^{\otimes d}& \gate{U_\Phi(x')}\qwbundle{} & \swap{-1} &  &  &  &
            \end{quantikz}
            \end{adjustbox}
            \label{fig:Swap-Test}
        }
    \end{minipage}
    \caption{Quantum circuits for IQP-like feature map, inversion test and swap test}
    \label{fig:mainfig}
\end{figure*}
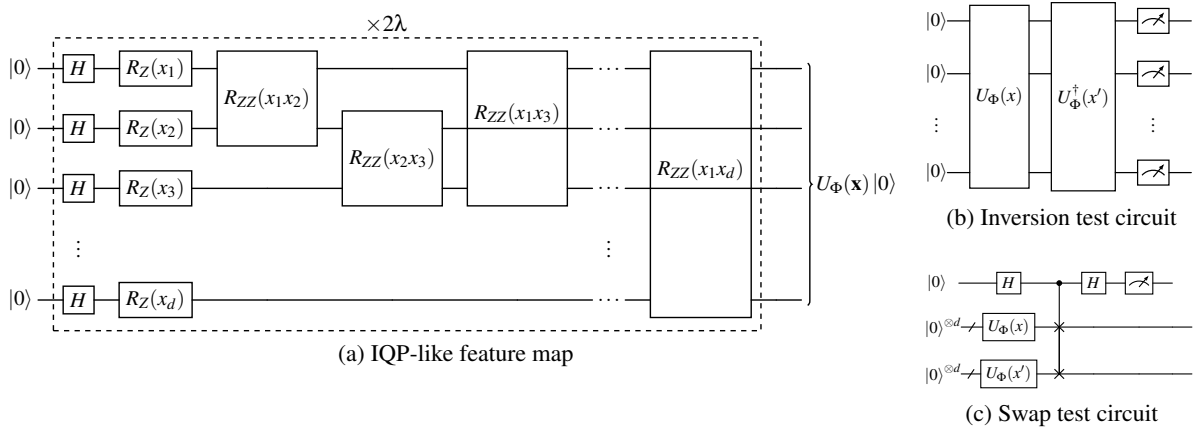

Quantum feature maps allow for the embedding of classical data into quantum states within Hilbert space $\mathcal{H}$ through the use of data-dependent quantum unitary gates $U_\Phi(x)$, mathematically expressed as $\ket{\Phi (x)} = U_\Phi(x) \ket{0}$. The IQP-like (Instantaneous Quantum Polynomial) feature map, recognized for being hard to simulate classically \cite{havlivcek2019supervised}, encodes a $d$-dimensional input $x_i$ into $d$ Qubits as follows:
\begin{equation}
    \small
    \begin{split}
    \ket{\Phi(x_i)} &= U_Z(x_i)H^{\otimes d}U_Z(x_i)H^{\otimes d} \ket{0^d},\\
    U_Z(x_i) &= \exp\left(\sum_{j=1}^{d} \lambda x_{ij}Z_j + \sum_{j=1}^d \sum_{j'=1}^d \lambda^2 x_{ij}x_{ij'} Z_j Z_{j'}\right),
    \end{split}
\end{equation}
where $\lambda$, influenced by data reuploading counts, impacts the kernel bandwidth analogously to $\gamma$ \cite{Shaydulin_2022}. We refer to Figure\autoref{fig:IQP-FeatureMap} for a visual representation. Upon data encoding, fidelity, distilled to the overlap of states for pure quantum states as $F(x, x') = \vert \braket{\Phi(x')}{\Phi(x)}\vert^2$, quantifies data similarity. 

Two prominent methods for measuring fidelity are the inversion and swap tests. The inversion test, detailed in Figure\autoref{fig:Inversion-Test}, calculates the overlap of two data points' pure quantum states with $O(n^2)$ kernel evaluations, but necessitates unitary feature maps and yields deeper circuits. Conversely, the swap test, depicted in Figure\autoref{fig:Swap-Test}, is applicable to both pure and mixed states but necessitates wider circuits. 
It is based on the swap trick \cite{hubregtsen2022training}, which deduces the inner product from the tensor product of density matrices $\rho_i$ and $\rho_j$ utilizing a swap gate $\mathbb{S}$, expressed in \autoref{eq:swap-trick}:
\begin{align}\label{eq:swap-trick}
     \Tr(\rho_i\rho_j) = \Tr(\mathbb{S}\rho_i \otimes \rho_j).
\end{align}

\subsection{Time Complexity for Quantum Kernels}\label{sec:Efficient_kernel}

Quantum one-class Support Vector Machines provide a nuanced method for detection but are hindered by formidable time complexity, particularly given the existing operational frequencies of quantum hardware. The intrinsic quadratic time complexity of one-class SVMs is significantly amplified in a quantum computing context, requiring a substantial number of repetitions (shots) to accurately measure fidelity between points due to its probabilistic nature; typically, at least 1000 shots per circuit measurement are necessary to obtain replicable results. Training a quantum one-class SVM on a large dataset, exemplified by the 284,000 instances from the Credit Card dataset in our experiments, would theoretically necessitate approximately $4 \times 10^{10}$ kernel function value calculations to construct the symmetrical kernel matrix. This implies a staggering requirement of $4 \times 10^{13}$ shots, which, with a measurement rate of 5kHz~\cite{haug2021large}, equates to a minimum training time of 255 years using a swap or inversion test kernel.

Reducing the dataset may abbreviate training time but risks degrading algorithmic performance and stability due to decreased representativity of training samples. This diminution can result in less dependable support vectors and decision functions, jeopardizing the reliability and consistency of the one-class SVM, especially when training data substantially deviates from the overall distribution. The challenge also permeates inference times, as predicting a new point demands evaluating the kernel function against all training points, elongating detection times and hindering real-time applications like patient monitoring and fraud prevention.

To surmount these obstacles, optimization of kernel calculation algorithms and investigation into innovative quantum measurement techniques, which could minimize the requisite shots, are pivotal. Viable approaches may encompass adapting classical methods to minimize data needed for kernel matrix computations while preserving performance, employing clustering, and applying matrix decomposition and approximation methods to avoid evaluating the kernel across the entire training set.

\section{\uppercase{Related Work}} \label{sec:related-work}

\subsection{Quantum Anomaly Detection}
This work augments the methodology propounded in \cite{kyriienko2022unsupervised}, amalgamating the one-class SVM with a computationally intricate kernel based on the IQP feature map (\autoref{fig:IQP-FeatureMap}) to secure a 20\% enhancement in average precision vis-à-vis their classical benchmark. Subsequent experiments herein adhere to this protocol as a quantum yardstick, exploring two strategies to diminish the time complexity relative to data size.

Hybrid quantum-classical models manifest as a prominent archetype in anomaly detection research. For instance, \cite{sakhnenko2022hybrid} innovatively refines the auto-encoder (AE) hidden representation by interfacing a parameterized quantum circuit (PQC) with its bottleneck, segueing into an unsupervised model post-training by substituting the decoder with an isolation forest, and vetting performance across multifarious data sets and PQC architectures. Concurrently, \cite{herr2021anomaly} pioneers an adaptation of the classical AnoGAN \cite{schlegl2017unsupervised} by deploying a Wasserstein GAN, wherein the generator is substituted with a hybrid quantum-classical neural network, and subsequently trained via a variational algorithm.

Contrastingly, quantum annealing approaches, such as the QUBO-SVM presented in \cite{wang2022integrating}, reformulate the conventional SVM optimization predicament into a quadratic unconstrained binary optimization problem (QUBO) which is amenable to resolution via quantum annealing solvers. Although retaining the conventional SVM optimization problem, this methodology expedites accurate predictions through adept kernel function identification, thereby facilitating plausible real-time anomaly detection.

\cite{ray2022classical} explores hybrid ensembles, constructing an amalgamation of bagging and stacking ensembles from assorted quantum and classical components, each playing a pivotal role in the anomaly detection framework. The amalgamated quantum components encapsulate disparate variable quantum circuit architectures, kernel-based, and quantum annealing-based SVMs, while the classical constituents encompass logistic regression, graph convolutional neural networks, and light gradient boosting models. Despite superficial similarity to the variable subsampling utilized herein, it’s noteworthy that the latter method's adoption of varying sub-sample sizes uniquely addresses the OC-SVM’s parameterization dependence.

\subsection{Efficient Gram Matrix Evaluation}
Quantum kernel methods, pivotal for various quantum machine learning applications, grapple with notable computational demands in matrix evaluations. Approaches to mitigate this complexity are: (i) \emph{Quantum-Friendly Classical Methods}, reducing kernel matrix elements to evaluate, and (ii) \emph{Quantum Native Methods}, minimizing the shot requirements yet necessitating classical post-processing, albeit feasibly parallelizable or vectorizable.

Randomized measurement kernels, pioneered by \cite{haug2021large} and utilized with hardware-efficient feature maps, achieved an expedited kernel measurement while approximating Radial Basis Function (RBF) kernels, demonstrated via both synthetic and MNIST data. Conversely, the classical shadow method, proposed by \cite{Huang_2020}, employs a similar quantum protocol but diverges in classical post-processing to provide classical state snapshots via the inversion of a quantum channel, often attaining reduced error in predicting second Rényi Entropy.

Variable subsampling, introduced by \cite{aggarwal2015theoretical}, and its sophisticated counterpart, variable subsampling with rotated bagging, offer an efficient ensemble training approach, leveraging varied sample sizes and rotational orthogonal axis system projections respectively. These methods not only confer computational efficiency but also harness an adaptive ensemble model training strategy, tested efficaciously with algorithms like local outlier factor (LOF) models and the k-NN algorithm.

The Distance-based Importance Sampling and Clustering (DISC) approach, by \cite{hajibabaee2021kernel}, and Block Basis Factorization, by \cite{wang2019block}, represent variants of matrix decomposition based methods for kernel approximation. While DISC employs cluster centroids as landmarks to formulate approximation matrices and assumes kernel matrix symmetry, block basis factorization avails randomized spectral value decomposition on cluster samples and foments a smaller, computationally tractable inner similarity matrix, demonstrating superior performance relative to the k-means Nyström method.

\section{\uppercase{Approaches}} \label{sec:approach}
In the preceding review of related work, several methodologies of efficient gram matrix evaluation were highlighted. However, moving forward, we will focus on two specific approaches which can be applied to both symmetric training kernel matrices and asymmetric ones used during prediction. Although the classical shadows and block basis factorization methods satisfy this criterion, we have decided to explore randomized measurements and variable subsampling methods because of their intuitive conceptual frameworks.


\subsection{Randomized Measurements Kernel}

Expanding upon the method of randomized measurements, suggested for future exploration in \cite{kyriienko2022unsupervised}, and practically employed in kernel calculation for classification by \cite{haug2021large}, this method endeavors to adeptly meld linear and quadratic complexities concerning data size. This is achieved respectively through acquiring measurements of feature maps and subsequent classical post-processing. The method notably diminishes the requisite quantum shots, thereby alleviating the overall computational burden.

The inception of the fidelity calculation was motivated by the possibility of conceiving the swap operator $\mathbb S$ as a quantum twirling channel $\Phi_N^{(2)}$ \cite{elben2019statistical}. 
Quantum twirling channels are operations commonly used in error correction. A 2-fold local quantum twirling channel applied to an arbitrary operator $O$ is articulated by
\begin{align}\label{eq:2-fold-quantum-twirl}
    \Phi_N^{(2)}(O) = \overline{(U^{\otimes 2})^\dagger O U^{\otimes 2}},
\end{align}
with $\overline{\mbox{\dots}\raisebox{2mm}{}}$ denoting the average over the unitaries \(U = \bigotimes_k^N U_k\) and \(U_k\) sampled from a unitary 2-design. A unitary t-design approximates property averages over all possible unitaries using a finite set, while the Haar measure provides a uniform sampling mechanism across unitary matrices.
\cite{elben2019statistical} demonstrate that the expectation value of applying such a twirling channel on an operator $O$ acting on two copies of a quantum state $\rho$ is obtainable from the probability products $P_U$ resulting from the measurements:
\begin{align}\label{eq:2nd-cross-correl}
    \sum_{s,s'} O_{s,s'} \overline{P_U(s)P_U(s')} = \Tr(\Phi_N^{(2)}(O) \;\rho \otimes \rho)
\end{align}
The coefficients $O_{s,s'}$  specific to the swap operator are derived by employing Weingarten calculus of Haar random unitaries and Schur-Weyl duality \cite{Roberts_2017} to calculate the purity of the state $\rho$, yielding:
\begin{align}
    O_{s,s'} = d^N(-d)^{D_(s,s')}.
\end{align}
A formula for quantum fidelities in terms of randomized measurement probabilities is obtained by utilizing the swap trick, then employing \autoref{eq:2-fold-quantum-twirl} with the coefficients from \autoref{eq:2nd-cross-correl}.

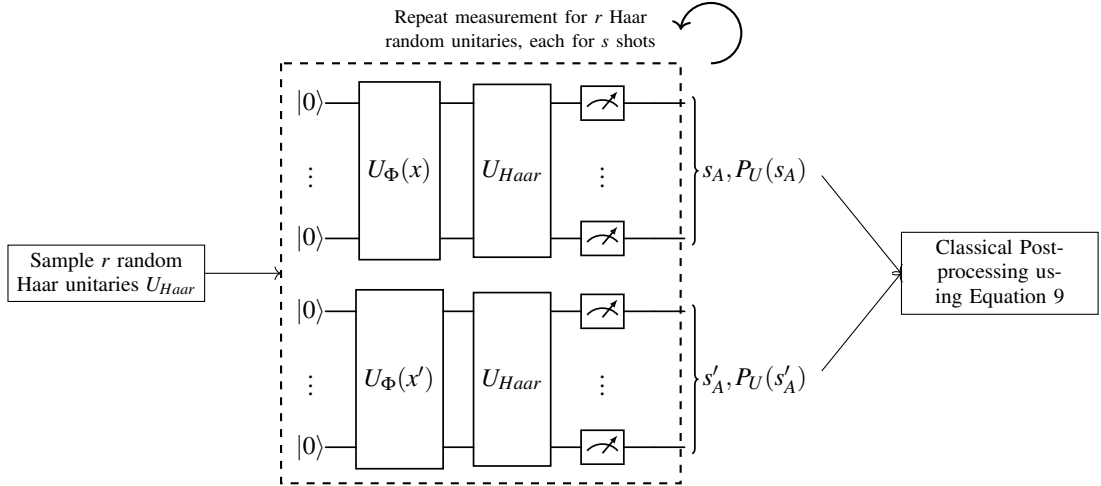
\begin{figure*}[htbp]
    \centering
        \begin{tikzpicture} 
        \node(A)[scale=0.8, draw=black,text width=3cm,align=center]{Sample $r$ random Haar unitaries $U_{Haar}$};
        \node(B)[scale=0.8, right=of $(A.east)$]{
        \begin{quantikz}[transparent,font=\large]
        \ket{0} & \gate[3]{U_\Phi(x)} & \gate[3]{U_{Haar}}&\meter{} &  & \rstick[3]{$s_A, P_U(s_A)$} \\
        \setwiretype{n}\vdots & &  & \vdots &  & \\
        \ket{0} &  & &\meter{} &  &\\
        \ket{0} & \gate[3]{U_\Phi(x')} & \gate[3]{U_{Haar}}&\meter{} &  & \rstick[3]{$s'_A, P_U(s'_A)$} \\
        \setwiretype{n}\vdots & &  & \vdots &  &  \\
        \ket{0} &  & &\meter{} &  &
        \end{quantikz}
        }; 
        \node(C)[scale=0.8, draw=black, right=of $(B)$,text width=3cm,align=center, xshift=4.5cm]{Classical Postprocessing using \autoref{eq:qRM-kernel}};
        \draw[->] (A.east) -- (B.west);
        \draw[->, sloped] (B.20) -- (C.west);
        \draw[->, sloped] (B.-20) -- (C.west);
        \draw[thick,dashed]($(B.north west)$) rectangle ($(B.south)+(1.7cm,0)$);
        \draw[thick, ->] (B.53) arc (-90:174:4mm) node[scale=0.8,left,text width=5cm,align=center]{\footnotesize Repeat measurement for $r$ Haar random unitaries, each for $s$ shots};;
        \end{tikzpicture}
    \tiny\caption{The protocol and the circuit architecture for calculating quantum kernel functions using randomized measurement.}
    \label{fig:Randomized-Measurements}
\end{figure*}

Local Haar random unitaries $U_\text{Haar}$ are constructed by tensoring sampled unitary $U_k \in SU(2)$ for each qubit. As illustrated in \autoref{fig:Randomized-Measurements}, each quantum circuit leverages a unitary $U_\Phi$, corresponding to the quantum feature map, and one of the $r$ different local Haar random unitaries $U_\text{Haar}$. Every circuit requires $s$ different shots. This yields $r$ sets of strings $s_A$ and their measurement probabilities $P_U^{(i)}(s_A)$ for the different random basis rotations $U_\text{Haar}$. 
Post-processing harnesses the formula:
\begin{equation}\label{eq:qRM-kernel}
    \begin{split}
        K(x_i,x_j) &= \text{Tr}(\rho_i\rho_j) \\
                   &= 2^d \sum_{s_A, s_A'} (-2)^{-D(s_A, s_A')} \overline{P^{(i)}_{U}(s_A)P^{(j)}_{U}(s_A')}
    \end{split}
\end{equation}

The statistical error in fidelity measurement, quantified as $\Delta G \approx \frac{1}{s\sqrt{r}}$, necessitates error mitigation, particularly pertinent in noisy hardware scenarios. A facilitative aspect of the randomized measurements approach is its provision for straightforward error mitigation, involving purities calculated and recorded in the diagonal of the training kernel matrix and exploiting a minor computational overhead in the testing phase for the asymmetrical kernel matrix.

The quantum kernel calculation segment, employing randomized measurements, presents a time complexity of $nrs$ ($n$: data size, $r$: basis rotation unitaries, $s$: shots per rotation), while classical post-processing demands $n^2$ complexity, albeit with an unfavorable exponential time complexity concerning the number of qubits (features). Implementation aligning with \cite{haug2021large} is accessible via the \texttt{Large Scale QML}\footnote{\url{https://github.com/chris-n-self/large-scale-qml}} GitHub repository, whereas randomized measurement processing and combination utilize functionalities from the \texttt{qc\_optim}\footnote{\url{https://github.com/chris-n-self/qc_optim}} repository. Accommodating the IQP-like feature and enabling interim kernel copy retention and calculation time-keeping mandated the development of a novel implementation.

\subsection{Variable Subsampling}

Introduced by \cite{aggarwal2017introduction}, variable subsampling addresses sensitivities in the one-class SVM to kernel choice and hyperparameter $\nu$ values by exploiting ensemble methods. Unlike bagging ensembles, variable sampling ensembles, while utilizing random subset selection for model training, employ varying sub-sample sizes, permitting sampling over parameter spaces, particularly those related to data size, like the expected anomaly ratio $\nu$ in the one-class SVM.

Supposing a variable subsampling ensemble comprises 3 OC-SVMs trained with $\nu=0.1$ and sample sizes of 53, 104, and 230, differing support vector lower bounds among components yield varied decision boundaries. These can subsequently be combined in a manner that diminishes the bias or variance in predictions. 

Ensemble construction commences with uniformly sampling $c$ different subsample sizes between 50 and 1000. Data subsets, corresponding to sampled sizes, are randomly selected from the dataset and employed to train base model versions, here, the quantum one-class SVM with inversion test. Although subsamples may contain identical elements, each avoids reusing data points to better reduce variance. Predictions are calculated by combining normalized (to zero mean and unit standard deviation) decision functions of all components, considering each is trained with distinct data sizes, thus possessing varied decision function value ranges \cite{aggarwal2015theoretical}. Simple averaging of outlier scores is advantageous for reducing variance and superior performance with smaller datasets. Applying the maximum score, conversely, curtails bias but elevates variance. Post decision function value extraction, the threshold function $\text{sgn}(.)$ derives the class label.

More components and a higher maximum subsample size facilitate optimal variance reduction, at the expense of increased computational resource and time demands. However, this trade-off can be managed through wise hyperparameter selection. For example, we choose a maximum subsample size of 100, instead of the recommended 1000 points from the original research, and use \(\lfloor \frac {n}{100}\rfloor\) components instead of 100. While this may negatively affect performance, it provides insight into the ensemble's behavior when scalability takes precedence.

The training phase exhibits approximately $c \cdot (\frac{n_\text{min} + n_\text{max}}{2})^2$ time complexity, where $c$ is the number of ensemble components, and $n_\text{min} = 50$ and $n_\text{max}$ are the minimum and maximum subsample sizes respectively. Incorporating scalability modifications, this becomes $\lfloor \frac{n}{100} \rfloor \cdot \left(\frac{50 + 100}{2}\right)^2$, indicating linear complexity. Testing similarly maintains linear time complexity, formulated as $c \cdot \frac{n_\text{min} + n_\text{max}}{2} \cdot n_\text{test}$, with $n_\text{test}$ representing test samples.

\section{\uppercase{Experimental Setup}} \label{sec:experimental-setup}

Ensuring the reproducibility of our experiments, this section meticulously delineates implementation aspects, encompassing pre-processing, various kernel calculation methodologies, and hyperparameter selection. Two distinct experiment sets facilitate the comparative analysis of our approaches:

\begin{enumerate}
    \item The first set aspires to examine performance, training, and testing durations with respect to data size. Primarily, it endeavors to contrast the computational efficiencies afforded by our methodologies, delineated in \autoref{sec:approach}, and to comprehend model responses to elevated data volumes.
    \item Focusing on the relationship between performance, computational time, and the number of features (or qubits), the second experiment set seeks not only to discern whether the methods impose detrimental effects on performance but also to elucidate the time feature/qubit relationship.
\end{enumerate}

Experiments are conducted using 15 distinct seeds, ranging from 0 to 14.




\subsection{Datasets}

Our Experiments utilize two datasets, distinguished by their synthetic or real-world origin, to investigate various methodologies. The synthetic dataset is employed exclusively in the first experiment set, given its two-dimensional nature and the associated limitations in exploring numerous features.

\subsubsection{Synthetic Data} 
Our synthetic dataset is a bi-dimensional, non-linearly separable dataset, derived by modifying an SKlearn OC-SVM demonstration\footnote{\url{https://scikit-learn.org/stable/auto_examples/svm/plot_oneclass.html}} to yield training samples of diverse sizes. Testing samples consistently comprise 125 points, incorporating a 0.3 anomaly ratio.

\subsubsection{Credit Card Fraud Data} 
The credit card fraud data\footnote{\url{https://www.kaggle.com/datasets/mlg-ulb/creditcardfraud}} encompasses approximately 284,000 datapoints, with 492 classified as anomalous (class 1), across 31 features. Omitting 'time' and 'amount', 28 PCA-applied, anonymized numerical features are retained. 
Each seed corresponds to a unique data split. The size varies for the set of experiments exploring the effects of data size, while it is kept at a constant 500 data samples for the set of experiments exploring the effects of the qubit/feature number.
Training data only contains non-anomaly samples while the test set always contains 125 points, which include 6 anomalies, achieving a 0.05 anomaly ratio.

\subsection{Data Pre-processing}

Distinctive pre-processing methodologies were necessitated based on the quantum kernel measurement technique applied and the data type (synthetic or real). 

\paragraph{Radial Basis Function (RBF) Kernel:} For implementations utilizing the RBF kernel, standard scaling was executed post data partitioning into training and test subsets, ensuring zero mean and unit standard deviation across all features. Subsequently, Principal Component Analysis (PCA) was employed to consolidate the data into the requisite number of features.

\paragraph{Inversion Test Kernels:} The application of inversion test kernels warranted an additional step following the pre-processing used for the RBF kernel. Given that data was utilized as rotation angles within the quantum circuit, a scaling by a factor of 0.1 was an imperative post-PCA application.

\paragraph{Randomized Measurements Kernels:} Adhering to guidelines by \cite{haug2021large}, the randomized measurements kernels necessitated a unique rescaling approach. Post-PCA, a second round of standard scaling was administered, succeeded by an additional rescaling using factor $\frac{1}{\sqrt{M}}$, with $M$ representing the post-PCA data dimensionality.

Note that while real data pre-processing was adapted in accordance with the specific quantum kernel measurement technique deployed, synthetic data was subjected to pre-processing solely when implementing the randomized measurement kernel, only requiring a step of standard scaling with an additional rescaling by factor $\frac{1}{\sqrt{M}}$, .


\subsection{Baselines}

For the real dataset, we aim to replicate the one-class SVM results from \cite{kyriienko2022unsupervised}, adopting them as quantum and classical benchmarks. Lacking detailed insight into their sampling method and explicit test set sizes, we employ random sampling to generate data sets comprising 500 training and 125 test data points, training the OC-SVM solely on non-anomaly data, like the original authors. An anomaly ratio of 0.05 is maintained in the test set.

The experiments concerning model responses to various data sizes retain these architectures, holding the number of PCA features steady while manipulating data sizes. Specifically, we utilize $2$ features for synthetic and $6$ for real data sets, exploring data sizes \( n \in \{250, 500, 750, 1000, 1250, 1500\} \).




\subsection{Models and Parameter Selection}

Employing both classical and quantum versions of the OC-SVM, we utilize \texttt{OneClassSVM} from the \texttt{SKlearn}\footnote{\url{https://scikit-learn.org/stable/modules/generated/sklearn.svm.OneClassSVM.html}} library. All classical models adopt the RBF kernel with $\gamma = \frac{1}{N\cdot \text{Var}(M)}$ and a consistent $\nu = 0.1$. 

Quantum circuits, pivotal for kernel calculations, are crafted with the \texttt{qiskit}\footnote{\url{https://qiskit.org/}} library, simulated via \texttt{qiskit\_aer.QasmSimulator}. Employing $\lambda = 3$ data reuploadings for all quantum methods' feature maps, inversion test kernel elements are determined using 1000 shots each. 

Randomized measurement circuits, realized with $r=30$ measurement settings and subjected to $s=9000$ shots each, leverage crucially implemented classical post-processing, which emphasizes minimal embedded loops and prioritizes efficient matrix operations. The variable subsampling method, utilizing $c=\lfloor\frac{n}{100}\rfloor$ components ($n$ representing the training set size) and a subsample size $n_s \in [50, 100]$ to ensure scalable model performance, invokes the inversion test for kernel calculations with 1000 shots per element. Employing a consistent $\nu$ value, the desired kernel matrix calculation function is passed as a callable kernel parameter, designed to accept two data sets and method-specific hyper-parameters, subsequently returning the computed kernel matrix.

\subsection{Performance Metrics for Imbalanced Data Sets}\label{subsec:PerformanceMetrics}
Analyzing highly imbalanced data sets, prevalent in anomaly detection, necessitates alternative metrics to accuracy due to its incapability to reflect model performance accurately across classes \cite{aggarwal2017introduction}.
Hence, derivatives of precision and recall like the F1 score and average precision gain prominence.



\paragraph{F1 Score} represents the harmonic mean of precision and recall, providing a balanced perspective on model performance regarding false positives and negatives. It is defined as
\begin{align}
    \text{F1 Score} = 2\cdot \frac{\text{Precision}\cdot\text{Recall}}{\text{Precision}+\text{Recall}}.
\end{align}

\paragraph{Average Precision} quantifies the model's capability to discern anomalies, irrespective of threshold, by measuring the area under the precision-recall curve. Specifically,
\begin{align}
    \text{AP} = \sum_k [\text{Recall}(k) - \text{Recall}(k+1)] \cdot \text{Precision}(k).
\end{align}

\noindent An average precision equal to the data's anomaly ratio signifies a model with no anomaly detection capability, while a score of 1 indicates a flawless detector. 
Consequently, our analysis prioritizes average precision, complemented by precision, recall, and F1 score insights.

\section{\uppercase{Results}} \label{sec:results}

\subsection{Performance Analysis Synthetic Dataset}\label{sec:perf-data-synth}


\begin{figure*}[htbp]
    \centering
    \includegraphics[width=0.93\linewidth]{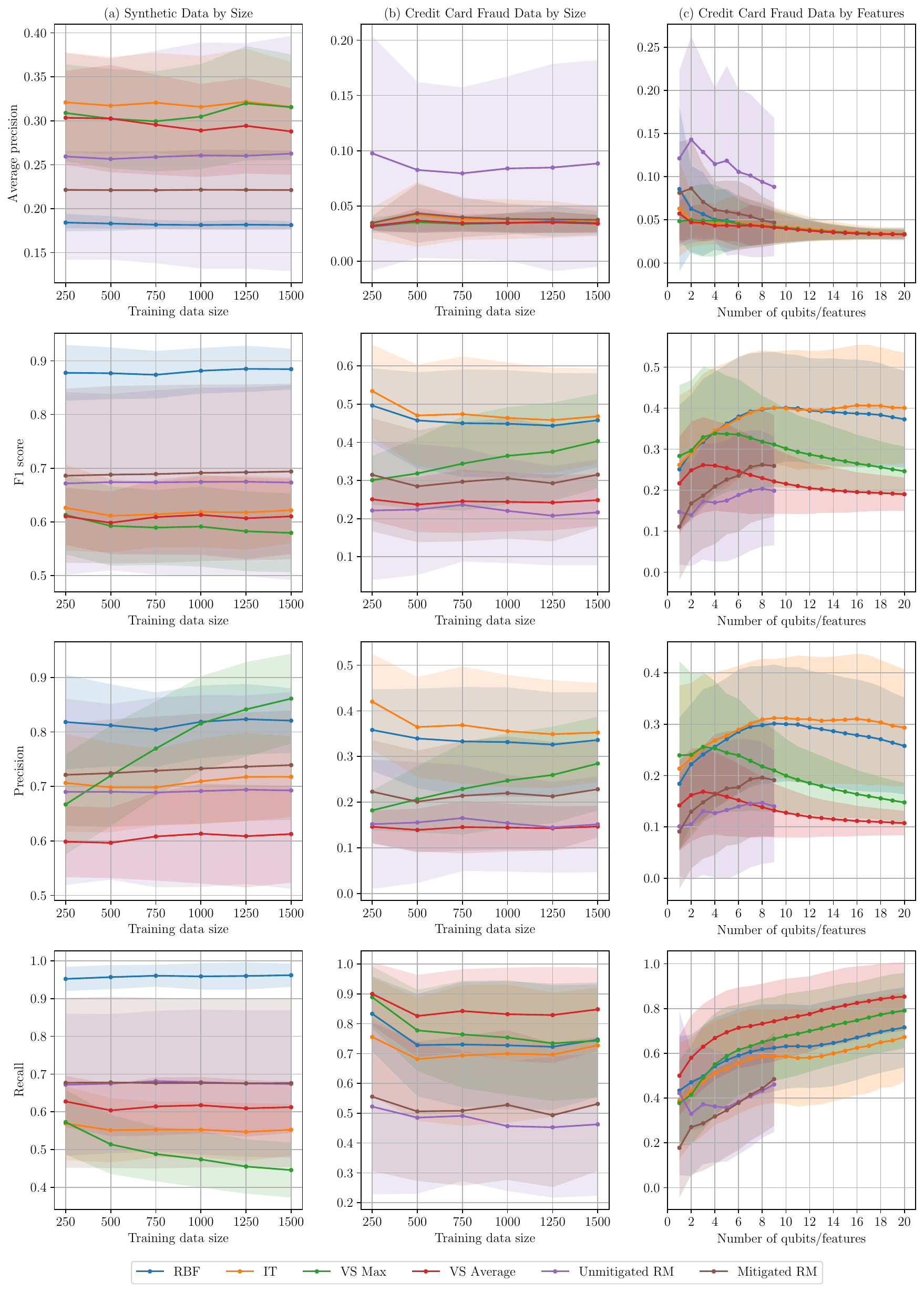}
    \caption{The performance of the models trained using the different methods. The first column of figure corresponds to the synthetic dataset, used with 2 features and various data sizes. The second and third columns correspond to the Credit Card dataset, with 6 as the number of features using various data sizes, then used with various numbers of features and a constant data size of 500.}
    \label{fig:performance}
\end{figure*}

The performance results utilizing synthetic data are presented in the first column of \autoref{fig:performance}a. Quantum methods demonstrate superior average precision compared to classical Radial Basis Function (RBF) models with the synthetic data. Despite this, given the anomaly ratio of 0.3, most methods underperform relative to a random detector, with inversion test and variable subsampling (using the maximum score) as exceptions. Variable subsampling not only surpasses randomized measurements in average precision but also exhibits comparable variance in ensemble results across different combination functions. Notably, it presents augmented results with the maximum over the average function for data sizes exceeding 750 points.

Upon applying the sign as a threshold, classical models outperform quantum models with the RBF kernel models securing an approximate F1 score of 0.88 and demonstrating lower variance. The randomized measurements and inversion test models follow in performance, with versions of the variable subsampling ensemble trailing. Despite the mitigated version of randomized measurements models achieving superior precision, F1 scores, and recall values were alike between mitigated and unmitigated versions. Contrarily, variable subsampling employing the maximum score delivers superior precision but inferior recall compared to its average score counterpart, suggesting its unsuitability for anomaly detection using the current threshold. 

Although results might appear conflicting, disparities may stem from suboptimal threshold selection for quantum methods. Given that average precision operates independently of a threshold, findings might imply an optimal threshold where quantum methods supersede classical RBF.

\subsection{Training and Testing Durations Synthetic Dataset}

\begin{figure*}[!ht]
    \centering
    \includegraphics[width=\linewidth]{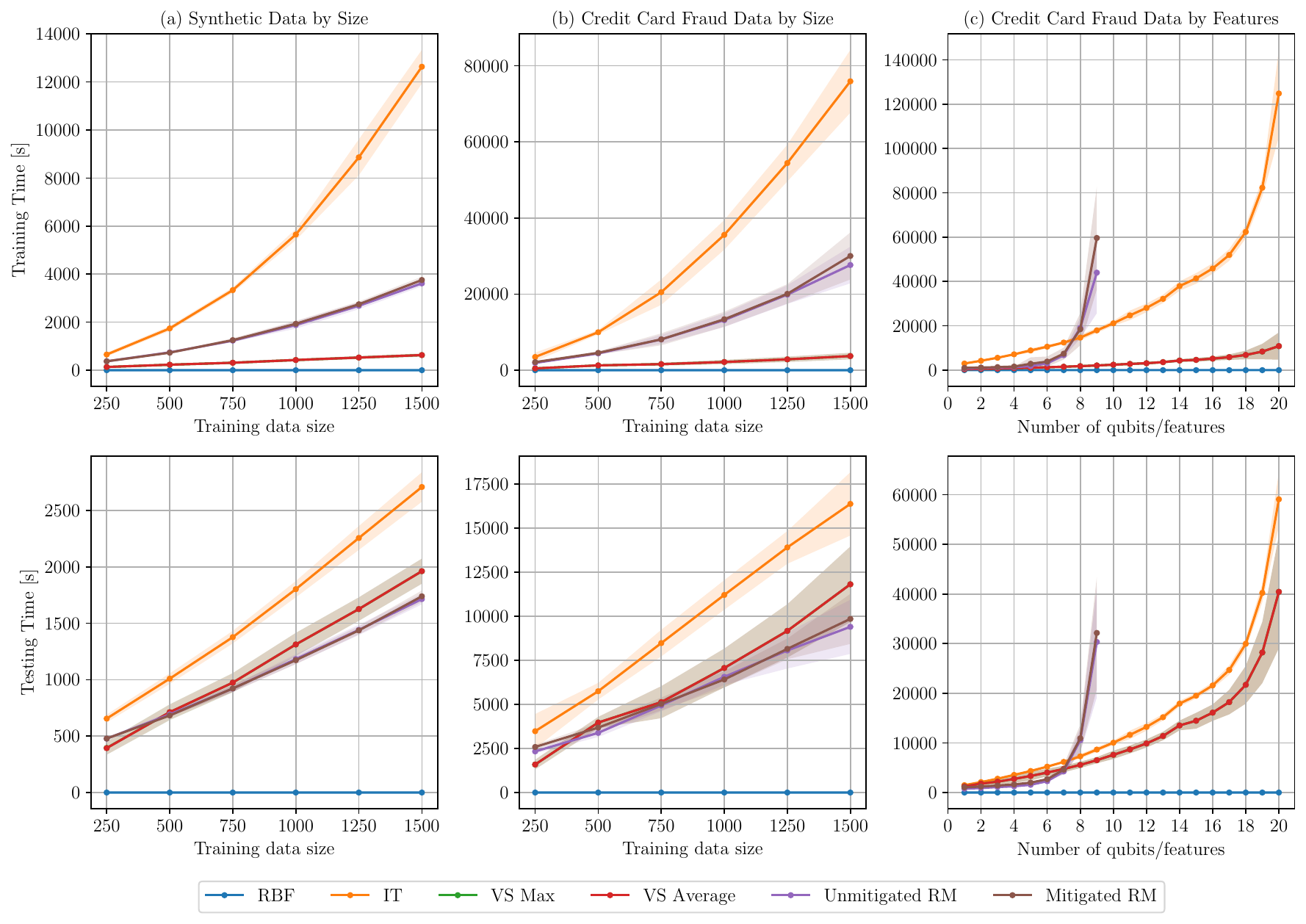}
    \caption{Training and testing durations in seconds, based on the data size.}
    \label{fig:time}
\end{figure*}
\autoref{fig:time}a depicts the notable efficiency of the variable subsampling method in reducing training durations, achieving the lowest among all employed quantum methods for both VS Max and VS Average. Interestingly, varied score functions do not impact the time consumption. Variable subsampling not only realizes linear time but also accomplishes a $\sim 95\%$ reduction in training time with 1500 data points, suggesting potential similar results in significantly less time given their average precision closely mirroring that of an individual one-class SVM utilizing the inversion test (\autoref{sec:preliminaries}). Furthermore, a $\sim 25\%$ enhancement in testing time is observed at 1500 training data points.

Mitigated and unmitigated randomized measurements methods yielded analogous times, corroborating that error mitigation imposes minimal overhead. While randomized measurements facilitate substantial training time reduction, complexity is non-linear relative to data size, attributed to the inclusion of classical post-processing in training time, which demands quadratic time complexity. Albeit the training time reduction from the randomized measurement method falls below that of variable subsampling, it yet provides $\sim 35\%$ and $\sim 12\%$ reduced testing times compared to the inversion test and variable subsampling respectively at 1500 training data points.

\subsection{Performance Analysis Relative To Data Size Credit Card Data}

\autoref{fig:performance}b reveals a predominant equivalence to a random detector across most models, with the unmitigated randomized measurements kernel being a notable exception. Despite this hinting at a potential performance threshold, the model's notable variance signifies an inherent instability.

At the utilized threshold, the inversion test and classical RBF exhibit comparable F1 scores in \autoref{fig:performance}b, succeeded by variable subsampling models employing maximum scoring, which seem to exhibit an enhancement with increased data usage. Despite affording the highest recall, both variable subsampling variations grapple with suboptimal precision.

Meanwhile, both randomized measurement kernels, under the current threshold, struggle with notably low recall, undermining their utility for anomaly detection.




\subsection{Time Complexity Relative Data Size Credit Card Data}
In \autoref{fig:time}b, we demonstrate that, analogous to synthetic data outcomes, employing variable subsampling assures linear time complexity and substantially curtails training time. Furthermore, randomized measurements realize up to a 50\% reduction in training time, even while tripling data utilization. Both variable sampling and randomized measurements maintain comparable, and notably reduced, inference times relative to the inversion test method.


\subsection{Performance Analysis Concerning Qubit Number Credit Card Fraud Dataset}

In our endeavor to replicate the results from \cite{kyriienko2022unsupervised}, we also explore additional metrics including precision, recall, and F1\_score, employing thresholds in model evaluation. The original study lacks explicit data selection methodology for training and testing, along with absent details on the number of runs, prompting us to opt for a uniformly random data selection.

Our findings, notably divergent from \cite{kyriienko2022unsupervised}, are presented in \autoref{fig:performance}c.

Average Precision reveals considerably less favorable outcomes for classical RBF and quantum inversion test than the aforementioned study, particularly over 15 different runs, underperforming even a random detector. Increasing feature/Qubit usage evens the average precision of the inversion test kernel, utilizing the IQP-like kernel, to that of the classical RBF, suggesting the original paper’s results might hinge more on beneficial sample selection for the quantum IQP-like kernel rather than demonstrating a clear quantum advantage. Notably, models exhibit enhanced stability—diminishing average precision variance—with augmented feature/Qubit use.

Thresholded metrics indicate marginal improvement with inversion tests over classical methods, yet an akin F1 score is attainable via the classical kernel with fewer Qubits. Variable subsampling, while offering the highest recall and consistent results, underachieves in F1\_score due to suboptimal precision—potentially attributable to conservative hyperparameter selections. An exploration into its performance with increased components and a larger maximum subsample could be interesting.

\subsection{Time Complexity Concerning Qubit Number Credit Card Data}
\autoref{fig:time}c illustrates a quadratic relationship between the training time and the number of Qubits for the inversion test. Despite a similar relationship, variable subsampling effectively mitigates training time, potentially attributed to its ensemble components utilizing the inversion test independently of Qubit number, constrained by limited and non-correlated component and data sizes. Conversely, the randomized measurements method sees escalated training times with 8 or more Qubits, hindering its utility for high-dimensional data sets. A rise in testing times is also observed with increasing Qubit numbers.


\section{Conclusion} \label{sec:conclusion}
This work explores two methodologies aimed at enhancing the scalability of the quantum one-class SVM in a semi-supervised framework: (1) the randomized measurements and (2) variable subsampling ensemble methods. The former, inherently quantum, realizes linear quantum time complexity but demands a quadratically complex classical post-processing based on data size. It generates kernel matrices by combining measurements from quantum feature maps of the data, executed through a randomized scheme. The latter adopts an ensemble approach, training multiple base model instances on varied-sized data subsets.

Two experimental datasets, synthetic and Credit Card, were deployed, revealing a marginal average precision improvement over the classical RBF for all methods in the synthetic data. Discrepancies emerged between outcomes from the Credit Card Fraud dataset and those in \cite{kyriienko2022unsupervised}. While models utilizing the inversion test and variable subsampling approximated the RBF and were relatively stable, those yielding higher average precision, like unmitigated randomized measurements, exhibited instability.

Future research trajectories include exploring the integration of variable subsampling with randomized measurements kernels and employing randomized measurements or variable subsampling alongside alternative kernels, such as learnable ones. Incorporating importance sampling in selecting the Haar random unitaries \cite{Rath_2021}, as suggested by \cite{haug2021large}, could refine the randomized measurements method. The classical shadow method, due to its lower average error compared to randomized measurements, and the examination of Nyström and block basis factorization approximations of a quantum kernel, warrant further exploration. Finally, enhancing the quantum SVM's performance through the proposed variable sampling method, by utilizing more components and enlarging maximum subsample sizes, remains a viable avenue, given the resultant training and testing time reductions.
\section*{\uppercase{Acknowledgements}}
This research is part of the Munich Quantum Valley, which is supported by the Bavarian state government with funds from the Hightech Agenda Bayern Plus.

\bibliographystyle{apalike}
{\small
\bibliography{main}}


\end{document}